\documentclass[usenatbib,useAMS]{mn2e}
\usepackage{graphicx}

\usepackage{aas_macros}
\newcommand{\gta}{{\small\raisebox{-0.6ex}
{$\,\stackrel{\raisebox{-.2ex}{$\textstyle >$}}{\sim}\,$}}}
\newcommand{\lta}{{\small\raisebox{-0.6ex}
{$\,\stackrel{\raisebox{-.2ex}{$\textstyle <$}}{\sim}\,$}}}

\title[]{SDSS J150722.30+523039.8: a CV formed directly from a detached white dwarf/brown dwarf binary?}

\author[]{S.\,P.\ Littlefair$^{1}$, V.\,S.\, Dhillon$^{1}$, 
T.\,R.\, Marsh$^{2}$, B.\,T.\, G\"{a}nsicke$^{2}$,
 I.\, Baraffe${^3}$, \newauthor C.\,A.\, Watson$^{1}$ \\
$^1$Dept of Physics and Astronomy, University of Sheffield, S3 7RH, UK \\
$^2$Dept of Physics, University of Warwick, Coventry, CV4 7AL, UK\\
$^3$Ecole Normale Sup\'{e}rieure de Lyon, CRAL, 46 all\'{e}e d'Italie,
69364 Lyon Cedex 07, CNRS UMR 5574, Universit\'{e} de Lyon 1, France\\}

\date{\center{\Large Submitted for publication in the Monthly Notices of the
Royal Astronomical Society \\
\vspace{.5cm} \today}}

\begin{document}
\maketitle

\begin{abstract} 
We present high-speed, three-colour photometry of the eclipsing
cataclysmic variable SDSS J150722.30+523039.8 (hereafter SDSS
J1507). This system has an orbital period of 66.61 minutes, placing it
below the observed ``period minimum'' for cataclysmic variables. We
determine the system parameters via a parameterised model of the
eclipse fitted to the observed lightcurve by $\chi^2$
minimisation. We obtain a mass ratio of $q = 0.0623 \pm 0.0007$ and an
orbital inclination $i = 83^{\circ}.63 \pm 0^{\circ}.05$. The primary
mass is $M_{\rmn{w}} = 0.90\pm0.01 M_{\sun}$. The secondary mass and
radius are found to be $M_{\rmn{r}} =0.056\pm0.001 M_{\sun}$ and
$R_{\rmn{r}} = 0.096 \pm 0.001 R_{\sun}$ respectively. We find a
distance to the system of $160 \pm 10$\, pc. The secondary star in
SDSS J1507 has a mass substantially below the hydrogen burning limit,
making it the second confirmed sub-stellar donor in a cataclysmic
variable.

The very short orbital period of SDSS J1507 is readily explained if
the secondary star is nuclearly evolved, or if SDSS J1507 formed
directly from a detached white dwarf/brown dwarf binary. Given the 
lack of any visible contribution from the secondary star, the very low
secondary mass and the low He\,{\small I} $\lambda6678$/H$\alpha$
emission line ratio, we argue that SDSS J1507 probably formed directly from a
detached white dwarf/brown dwarf binary. If confirmed, SDSS J1507 will
be the first such system identified. The implications for binary star
evolution, the brown-dwarf desert and the common envelope phase are
discussed.
\end{abstract} 


\section{Introduction}
\label{sec:introduction}
Cataclysmic variable stars (CVs) are a class of interacting binary
system undergoing mass transfer, usually via a gas stream and accretion disc,
from a Roche-lobe filling secondary to a white dwarf primary. A bright
spot is formed at the intersection of the disc and gas stream, giving
rise to an `orbital hump' in the lightcurve at phases $0.6-1.0$ due
to foreshortening of the bright-spot. \citet{warner95a} gives a
comprehensive review of CVs. The lightcurves of eclipsing CVs can be
quite complex, with the accretion disc, white dwarf and bright-spot
all being eclipsed in rapid succession. With sufficient
time-resolution, however, this eclipse structure allows the system
parameters to be determined to a high degree of precision \citep{wood86a}.

SDSS J1507 was discovered as part of the systematic effort to catalog
and characterise the CVs found by the Sloan Digital Sky Survey
\citep{szkody05}.  Its spectrum shows a strong blue continuum and
absorption lines from the white dwarf, as well as the broad,
double-peaked emission lines from the accretion disc, which are
characteristic of high inclination systems \citep{horne86}.  Follow up
photometry of SDSS J1507 \citep{szkody05} showed deep eclipses, and an
orbital period of $\sim 67$ minutes. The orbital period of SDSS J1507
places it below the {\em period minimum}, a sharp cut-off in the
number of CVs at $\simeq 78$ mins. Whilst there are objects related to
CVs with orbital periods of less than 78 minutes, known as AM CVn
stars, they are believed to contain a white dwarf and a degenerate
helium white dwarf or a semi-degenerate helium secondary star
\citep{nelemans01}.  Only two other CVs exist with orbital periods
below the period minimum, V485 Cen \citep{augusteijn96} and EI Psc
\citep{thorstensen02}. Possible explanations for the presence of CVs
below the period minimum include a donor which has undergone
significant nuclear evolution in its core prior to mass transfer
\citep[e.g.][]{thorstensen02}, or the direct formation of a CV from a
detached white dwarf/brown dwarf binary \citep{politano04}. The
presence of deep eclipses in SDSS J1507 allows a precise determination
of the donor star's mass, which may in turn allow us to decide between
these two scenarios.
 
In this paper we present {\sc ultracam} $u'g'r'$ lightcurves of SDSS
J1507, and use these lightcurves to derive the system parameters. The
observations are described in section~\ref{sec:obs}, the results are
presented in section~\ref{sec:results}, and discussed in
section~\ref{sec:disc}.

\section{Observations}
\label{sec:obs}
\begin{table*}
\begin{center}
\caption[]{Journal of observations. All observing nights were clear of
  cloud, but the night of 03$^{rd}$ Mar 2006 suffered from high
  humidity, whilst the seeing was poor for the night of 05$^{th}$ Mar
  2006 . The dead-time between exposures was 0.027~s for all
  observations. The relative GPS timestamping on each data point is
  accurate to 50 $\mu$s.}
\begin{tabular}{crrcccccccc}
\hline
Date & Start Phase & End Phase  & Filters & Exposure time (s) & 
Data points & Eclipses & Seeing (arcsec) & Airmass \\ \hline
2006 Mar 03 & -0.50   &  0.22   & {\em u}$^{\prime}${\em g}$^{\prime}${\em
  r}$^{\prime}$ & 1.995 & 1450 & 1 & 1.1--1.5 & 1.092--1.096 \\
2006 Mar 04 & 17.57   & 18.18   & {\em u}$^{\prime}${\em g}$^{\prime}${\em
  r}$^{\prime}$ & 1.995 & 1206 & 1 & 1.4--2.8 & 1.441--1.630 \\
2006 Mar 04 & 19.83   & 20.14   & {\em u}$^{\prime}${\em g}$^{\prime}${\em
  r}$^{\prime}$ & 1.995 &  636 & 1 & 1.4--2.2 & 1.138--1.164 \\
2006 Mar 05 & 38.53   & 39.16   & {\em u}$^{\prime}${\em g}$^{\prime}${\em
  r}$^{\prime}$ & 1.995 & 1257 & 1 & 2.0--4.0 & 1.620--1.900 \\
2006 Mar 05 & 41.39   & 42.14   & {\em u}$^{\prime}${\em g}$^{\prime}${\em
  r}$^{\prime}$ & 1.995 & 1504 & 1 & 2.0--5.0 & 1.112--1.164 \\
2006 Mar 07 & 86.48   & 87.42   & {\em u}$^{\prime}${\em g}$^{\prime}${\em
  r}$^{\prime}$ & 1.995 & 1893 & 1 & 0.8--1.6 & 1.096--1.141 \\
2006 Mar 08 &103.72   &104.29   & {\em u}$^{\prime}${\em g}$^{\prime}${\em
  r}$^{\prime}$ & 1.995 & 1161 & 1 & 1.4--2.4 & 1.473--1.666 \\
2006 Mar 08 &104.91   &105.14   & {\em u}$^{\prime}${\em g}$^{\prime}${\em
  r}$^{\prime}$ & 1.995 &  454 & 1 & 1.0--1.5 & 1.286--1.327 \\
\hline
\end{tabular}
\label{journal}
\end{center}
\end{table*}

On nights between Mar 03$^{rd}$ 2006 and Mar 08$^{th}$ 2006, SDSS
J1507 was observed simultaneously in the SDSS-$u'g'r'$ colour bands
using {\sc ultracam} \citep{dhillon07} on the 4.2-m William Herschel
Telescope (WHT) on La Palma. A complete journal of observations is
shown in table~\ref{journal}. The observations were taken between
airmasses of 1.0--1.9, in typical seeing conditions of 1.5 arcsecs,
but with a range of 0.6--5.0 arcsecs.  The data were taken in
photometric conditions. Eight eclipses were observed in total. Data
reduction was carried out in a standard manner using the {\sc
ultracam} pipeline reduction software, as described in
\cite{feline04a}, and a nearby comparison star was used to correct the
data for transparency variations. Because of the absence of a
comparison star which was sufficiently bright in the $u'$-band, the
$u'$-band data was corrected using the $g'$-band data for the
comparison star, with appropriate corrections for extinction.

\section{Results}
\label{sec:results}

\subsection{Orbital Ephemeris}
\label{subsec:ephem}
\begin{table}
\begin{center}
\caption[]{Mid-eclipse timings (HJD + 2453798). The
uncertainties for the values measured from our eclipses are
8.0$\times10^{-6}$ for the red lightcurves and 4.0$\times10^{-6}$
for the green lightcurves.}
\begin{tabular}{rrr}
\hline
Cycle No. &  {\em g}$^{\prime}$ & {\em r}$^{\prime}$\\
\hline
0   &   0.738841&  0.738845\\
18  &   1.571497&  1.571508\\
20  &   1.664004&  1.664012\\
39  &   2.542924&  2.542920\\
42  &   2.681689&  2.681683\\
87  &   4.763315&  4.763321\\
104 &   5.549707&  5.549720\\
105 &   5.595994&  5.595964\\
\hline
\end{tabular}
\label{eclipse_times}
\end{center}
\end{table}
The times of white dwarf mid-ingress $T_{\rmn{wi}}$ and mid-egress
$T_{\rmn{we}}$ were determined by locating the minimum and maximum
times, respectively, of the lightcurve derivative.  Mid-eclipse
times, $T_{\rmn{mid}}$, were determined by assuming the white dwarf
eclipse to be symmetric around phase zero and taking
$T_{\rmn{mid}}=(T_{\rmn{we}}+T_{\rmn{wi}})/2$. Mid-eclipse timings
are shown in table~\ref{eclipse_times}. The errors on
our mid-eclipse times were adjusted to give $\chi^{2} = 1$. The $g'$-band
and the $r'$-band lightcurves only were used, given the significantly poorer
quality of the $u'$-band lightcurve. The results from the two lightcurves were
combined with a weighted mean,  giving an ephemeris of:

\begin{displaymath}
\begin{array}{ccrcrl}
\\ HJD & = & 2453798.738844 & + & 0.04625829 & E.  \\
 & & 2 & \pm & 3 &
\end{array} 
\end{displaymath}

There was no significant deiviation from linearity in the $O-C$
times. This ephemeris was used to phase our data for the analysis
which follows.

\subsection{Lightcurve morphology and variations}
\label{subsec:lcurve}
\begin{figure}
\begin{center}
\includegraphics[scale=0.35,trim=0 0 0 0,clip,angle=-90]{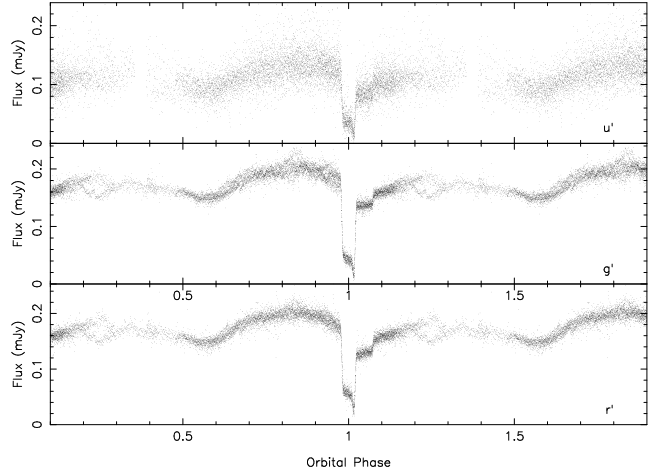} 
\caption{The phase folded $u'g'r'$ (from top to bottom) lightcurves
  of SDSS J1507. All eight observed eclipses are shown superimposed
  to emphasise the lack of variability between them.}
\label{fig:lcurve}
\end{center}
\end{figure}

Figure~\ref{fig:lcurve} shows the eight observed eclipses of SDSS
J1507, folded on the orbital phase. The white dwarf ingress and egress
features are clearly visible, as are the ingress and egress features
of the bright spot \citep[see chapter 2.6.2 of][for an illustrated
example of an eclipse in a typical dwarf nova system]{warner95a}. The
lightcurve is dominated by emission from the bright spot and the white
dwarf. The eclipse of the accretion disc is difficult to discern by
eye, implying that the accretion disc contributes little light to this
system. There is no indication of ellipsoidal variations from the
secondary star, but a double-humped morphology is seen. This
double-humped morphology likely arises from the bright spot being
visible throughout the orbit, which implies the accretion disc in this
system is optically thin. There is very little variation between the
individual eclipse lightcurves, implying that photometric precision
and flickering dominate over systematic errors in lightcurve
measurement for this system.

\subsection{A parameterised model of the eclipse}
\label{sec:model}
To determine the system parameters we used a physical model of the
binary system to calculate eclipse lightcurves for each of the various
components. \cite{feline04} showed that this method gives a more
robust determination of the system parameters in the presence of
flickering than the derivative method of \cite{wood86a}. We used the
technique developed by \citet{horne94} and described in detail
therein. This model relies on two critical assumptions: that the
bright spot lies on the ballistic trajectory from the secondary star,
and that the white dwarf is accurately described by a theoretical
mass-radius relation.  The error introduced by the latter assumption
can be estimated by comparing the results from different theoretical
models, and we show later in this section that it is smaller than the
statistical errors. Whilst the former assumption cannot be directly
tested, \cite{feline05} show that masses derived with this model are
consistent with other methods commonly employed in cataclysmic
variables over a wide range of orbital periods, including systems with
similar accretion discs to SDSS J1507.

\begin{figure}
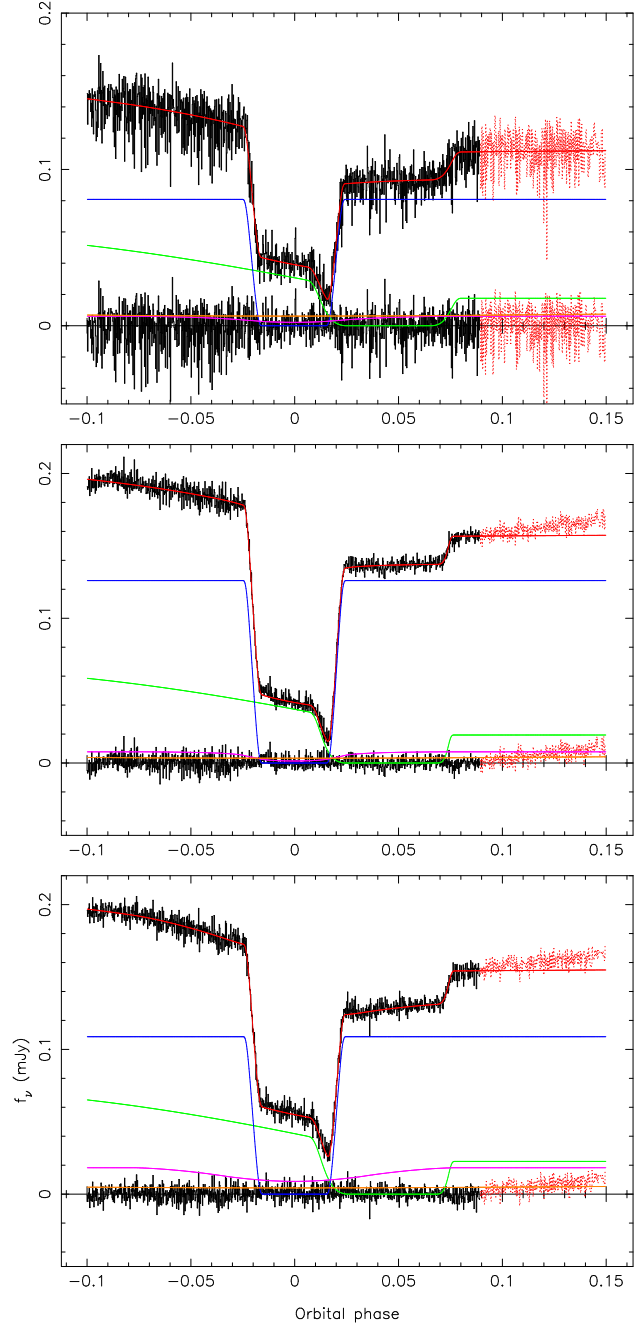

\begin{center}
\includegraphics[scale=0.35,angle=-90,trim=0 30 20 0,clip]{plots/blu.ps} 
\includegraphics[scale=0.35,angle=-90,trim=0 30 20 0,clip]{plots/grn.ps} 
\includegraphics[scale=0.35,angle=-90,trim=0 30 -20 0]{plots/red.ps} 
\caption{The phased folded $u'g'r'$ (from top to bottom) lightcurves
  of SDSS J1507, fitted separately using the model described in
  section~\ref{sec:model}. The data (black) are shown with the fit
  (red) overlaid and the residuals plotted below (black). Below are
  the separate lightcurves of the white dwarf (blue), bright spot
  (green), accretion disc (purple) and the secondary star
  (orange). Data points ignored in the fit are shown in red (light grey). }
\label{fig:model}
\end{center}
\end{figure}

We fit the model to all the observed eclipses, which were phase-folded
and binned into groups of three data points. The data and their model
fits are shown in figure~\ref{fig:model}. The 10 parameters that
control the shape of the lightcurve are as follows:
\begin{enumerate}
\item The mass ratio, $q=M_r/M_w$.
\item The white dwarf eclipse phase full-width at half-depth, $\Delta\phi$.
\item The outer disc radius, $R_{\rmn{d}}/a$.
\item The white dwarf limb-darkening coefficient, $U_{\rmn{w}}$.
\item The white dwarf radius, $R_{\rmn{w}}/a$.
\item The bright-spot scale, $S/a$. The bright-spot is modelled as a
linear strip passing through the intersection of the gas stream and
disc. The intensity distribution along this strip is given by
$(X/S)^{2}e^{-X/S}$, where $X$ is the distance along the strip.
\item The bright-spot tilt angle, $\theta_{\rmn{B}}$,
measured relative to the line joining the white dwarf and the
secondary star. This allows adjustment of the phase of the orbital hump.
\item The fraction of bright-spot light which is isotropic, $f_{iso}$.
\item The disc exponent, $b$, describing the power law of the radial
intensity distribution of the disc.
\item A phase offset, $\phi_{0}$.
\end{enumerate}

The {\sc amoeba} algorithm (downhill simplex; \citealt{press86}) was
used to adjust all parameters bar $U_W$ to find the best fit. A linear
regression was used to scale the four lightcurves (for the white
dwarf, bright-spot, accretion disc and secondary) to fit the observed
lightcurves in each passband. The excellent agreement between model
and data gives us confidence that our simple model accurately
describes the system. However, the model is a poorer fit to the data
after eclipse. This is most likely because the accretion disc in SDSS
1507+5230 is optically thin, allowing the bright spot to remain
visible when on the far side of the accretion disc. Such an effect is
not included in our simple model. To limit the effects that the poor
fit to post-eclipse data may have on the resulting system parameters
we excluded the regions shown in red (light grey) in
figure~\ref{fig:model} from the fit.  

In order to estimate the errors on each parameter once the best fit
had been found, we perturbed one parameter from its best fit value by
an arbitrary amount (initially 5 per~cent) and re-optimised the rest
of them (holding the parameter of interest, and any others originally
kept constant, fixed). We then used a bisection method to determine
the perturbation necessary to increase $\chi^{2}$ by $1$, i.e.\
$\chi^{2}-\chi_{\rmn{min}}^{2}=\Delta\chi^{2}=1$. The difference
between the perturbed and best-fit values of the parameter gave the
relevant $1\sigma$ error \citep*{lampton76}.  In the absence of
systematic errors, or flaws in the model's assumptions, the method
described above produces errors and parameters which are robust, and
unique \citep{littlefair06}. We are able to estimate the effects of
systematic errors in our photometry by fitting each of our eight
oberved eclipses separately, and comparing the spread of results with
the formal errors produced by the method above. Only the $g'$- and
$r'$-band lightcurves are of sufficient quality to do this. We find
that the spread in parameters estimated from each individual $r'$-band
eclipse are in good agreement with the formal errors, whilst the
spread in parameters estimated from the individual $g'$-band eclipses
suggest that the formal errors underestimate the true uncertainty in
some parameters, by up to a factor of two. The $g'$- and $r'$-band
lightcurves have almost identical signal/noise ratios, and so we might
expect them to suffer from systematic errors in the photometry at
roughly the same level. On the other hand, the $g'$-band lightcurve
shows much more flickering prior to eclipse (the RMS before eclipse is
about 3 times higher in $g'$ than $r'$). It is possible that the
presence of flickering affects the model estimation in a {\em
systematic} way. In this case, the formal error is an appropriate
estimate. Alternatively, the $g'$-band lightcurve might be more
affected by systematic errors in the photometry than the $r'$-band
lightcurve, in which case the formal errors underestimate the true
uncertainty by up to a factor of two. Since it is not clear which
error estimate is more reliable, in each case we adopt the largest of
the two estimates. These errors are shown in
table~\ref{parameters_lfit}.

The data were not good enough to determine the white dwarf
limb-darkening coefficient, $U_{\rmn{w}}$, accurately. To find an
appropriate limb-darkening coefficient, we obtained an estimate of the
effective temperature and mass of the white dwarf from a first
iteration of the method below, and assuming a limb-darkening
coefficient of 0.5. The mass and effective temperature were then used
in conjunction with the stellar atmosphere code of \cite{gaensicke95}
to generate angle-dependent white dwarf model spectra. To convert the
spectra to observed fluxes the model spectra were folded through
passbands corresponding to the instrumental response in each filter;
the effects of the SDSS filter set, the {\sc ultracam} CCD responses
and the dichroics used in the instrumental optics were taken into
account. These fluxes were then fit as a function of the limb position
in order to derive limb-darkening parameters for each band. The values
adopted are given in table~\ref{parameters_lfit}. A second iteration
using these values for the limb-darkening parameter gave the final
values for each parameter.  Comparison of the limb-darkening
parameters provided by different atmosphere models
\citep{gaensicke95,hubeny95} suggests an uncertainty in $U_{\rmn{w}}$
of $\sim5$ percent. This corresponds to an uncertainty in
$R_{\rmn{w}}/a$ of $\sim1$ percent, which was added in quadrature to
the error in $R_{\rmn{w}}/a$ provided by the fitting process.

\begin{table*}
\begin{center}
\caption[]{Parameters fitted using a modified version of the model of
\citet{horne94}. The fluxes of each component are also shown. Prior to
fitting, the data were phase-folded and binned by three data points.
Note that the orbital inclination $i$ is not a fit parameter but is
calculated using $q$ and $\Delta\phi$.}
\begin{tabular}{lccc}
\hline
Band & {\em u}$^{\prime}$ & {\em g}$^{\prime}$ & {\em r}$^{\prime}$ \\
\hline
Mass ratio $q$ & $0.060\pm0.002$ & $0.0630\pm0.001$ & $0.0622\pm0.001$ \\
Eclipse phase width $\Delta\phi$& $0.0405\pm0.0001$ & $0.04054\pm0.00005$ & $0.0403\pm0.0002$ \\
Outer disc radius $R_{\rmn{d}}/a$ & $0.343\pm0.007$ & $0.329\pm0.004$ & $0.330\pm0.004$\\
White dwarf limb darkening $U_{\rmn{w}}$ & $0.457\pm0.022$ & $0.346\pm0.017$ & $0.298\pm0.015$ \\
White dwarf radius $R_{\rmn{w}}/a$ & $0.0177\pm0.0008$ & $0.0169\pm0.0006$ & $0.0170\pm0.0005$ \\
bright-spot  scale $S/a$ & $0.040\pm0.008$ & $0.028\pm0.004$ & $0.032\pm0.004$\\
bright-spot  orientation $\theta_{\rmn{B}}$ & $161\fdg\pm2\fdg$ & $157\fdg\pm6\fdg$ & $157\fdg\pm6\fdg$ \\
Isotropic flux fraction $f_{iso}$ & $0.29\pm0.02$ & $0.30\pm0.06$ & $0.31\pm0.05$ \\  
Disc exponent $b$ & $0\pm20$ & $-2.0\pm2.0$ & $-0.4\pm0.5$ \\ 
Phase offset $\phi_{0}$ & $38\pm8\times10^{-5}$ & $15\pm4\times10^{-5}$ & $10\pm8\times10^{-5}$ \\
$\chi^{2}$ of fit & 2555 & 2108 &  1470 \\
Number of datapoints $\nu$ &  1269 &  1269 &  1269 \\
\hline
Flux (mJy) \\
\hspace{0.1cm} White dwarf    & $0.0809\pm0.0012$ & $0.1260\pm0.0004$ & $0.1088\pm0.0004$ \\
\hspace{0.1cm} Accretion disc & $0.0061\pm0.0022$ & $0.0077\pm0.0011$ & $0.0182\pm0.0010$ \\
\hspace{0.1cm} Secondary      & $0.0081\pm0.0013$ & $0.0048\pm0.0005$ & $0.0058\pm0.0007$ \\
\hspace{0.1cm} bright-spot    & $0.0593\pm0.0007$ & $0.0651\pm0.0002$ & $0.0072\pm0.0003$ \\
\hline
\end{tabular}
\label{parameters_lfit}
\end{center}
\end{table*}

We calculated the remaining system parameters following the method
described in \cite{littlefair06}. The model fitting provides estimates
of $q$, $\Delta\phi$ and $R_{\rmn{w}}/a$. The orbital inclination $i$
is determined from $q$ and $\Delta\phi$, using geometrical arguments
\citep{bailey79}.  Independent fits to the $r'$-, $g'$- and $u'$-band
lightcurves yield consistent values of $q$, $\Delta\phi$ and
$R_{\rmn{w}}/a$ - a weighted mean of these values was adopted for the
analysis that follows. The model also yields fluxes for the white
dwarf component, derived by fitting the size of the white dwarf
ingress and egress. The signal-to-noise ratio of our data is good, and
the white dwarf fluxes are likely to be dominated by systematic
errors.  We added systematic errors of 1 percent to our white dwarf
fluxes to account for this. A white dwarf temperature was found by
fitting the white dwarf colours to the predictions of white dwarf
model atmospheres \citep{bergeron95}.  A mass for the white dwarf can
then be derived from Kepler's 3$^{rd}$ Law, the orbital period, the
mass ratio and a mass-radius relationship for the white dwarf. We
adopted an appropriate mass-radius relationship for the white dwarf
\citep{wood95}, taking into account the effective temperature found
above. Comparison of different white dwarf models \citep{wood95,
bergeron95, panei00}, differing hydrogen envelope masses and
correction to different effective temperatures revealed that the
dominant source of uncertainty in the white dwarf mass is the
uncertainty in $R_{\rmn{w}}/a$. Once the white dwarf mass is known,
the mass of the donor star follows from $q$. The volume equivalent
radius of the donor star can be calculated, assuming the donor star
fills its Roche Lobe. A distance to the system was derived by
comparing the white dwarf fluxes in table~\ref{parameters_lfit}, and
the predicted fluxes from \cite{bergeron95}; the uncertainty in the
distance is dominated by the uncertainty in the white dwarf
temperature. The final adopted system parameters are shown in
table~\ref{parameters}.

\begin{table}
\begin{center}
\caption[]{System parameters of SDSS J1507 derived using mass--radius
  relation of \protect\cite{wood95}, at the appropriate
  $T_{\rmn{w}}$. $R_{\rmn{r}}$ is the volume radius of the secondary's
  Roche lobe \citep{eggleton83}.  The weighted means of the
  appropriate values from Table~\ref{parameters_lfit} are used for the
  system parameters. }
\begin{tabular}{lcccc}
\hline
\hline
Inclination $i$ & $83\fdg63 \pm 0\fdg05$\\
Mass ratio $q=M_{\rmn{r}}/M_{\rmn{w}}$ & $0.0623 \pm 0.0007$ \\
White dwarf mass $M_{\rmn{w}}/M_{\sun}$ & $0.90 \pm 0.01$ \\
Secondary mass $M_{\rmn{r}}/M_{\sun}$ & $0.056 \pm 0.001$ \\
White dwarf radius $R_{\rmn{w}}/R_{\sun}$ & $0.0091 \pm 0.0001$ \\
Secondary radius $R_{\rmn{r}}/R_{\sun}$ & $0.096 \pm 0.001$ \\
Separation $a/R_{\sun}$ & $0.535 \pm 0.002$ \\
White dwarf radial velocity $K_{\rmn{w}}/\rmn{km\;s^{-1}}$ & $34.1 \pm 0.4$ \\
Secondary radial velocity $K_{\rmn{r}}/\rmn{km\;s^{-1}}$ & $548 \pm 2.0$ \\
Outer disc radius $R_{\rmn{d}}/a$ & $0.333 \pm 0.003$ \\
White dwarf temperature $T_{\rmn{w}}/\rm{K}$ & $11000 \pm 500$ \\
Distance $d/\rmn{pc}$ & $160 \pm 10$ \\
\hline
\end{tabular}
\label{parameters}
\end{center}
\end{table}

\section{Discussion}
\label{sec:disc}

\subsection{The evolutionary status of SDSS J1507}
\label{subsec:evol}

The orbital period distribution of CVs shows a sharp cut-off around
$\sim 80$ mins, known as the {\em period minimum} \citep[see figure 1
of][for example]{knigge07}. The existence of a period minimum is
easily explained in terms of the response of the secondary star to
mass loss \citep{paczynski81}. Above the period minimum, the secondary
star shrinks in response to mass loss and the orbital period
decreases. As the secondary star nears the substellar limit, however,
the thermal time-scale becomes longer than the mass-transfer
timescale. When this happens the secondary is not able to shrink fast
enough and the period begins to increase. A long-standing problem with
this theory is that theoretical models persistently predict a period
minimum around 67 minutes \citep[e.g.][]{kolb99}, about 10 minutes short of
the observed period minimum. With an orbital period of 67 minutes and
a brown dwarf donor, SDSS J1507 could represent the ``true'' period
minimum for CVs. Indeed, the secondary star mass and radius presented
in table~\ref{parameters} are consistent with the predicted mass and
radius for a system near the {\em predicted} minimum orbital period
\citep{kolb99}.  There is one major problem with this scenario
however; where are all the systems between 67 minutes and the observed
period minimum at $\sim 80$ minutes? Population synthesis models
predict a large number of systems near the period minimum
\citep[e.g][]{kolb93}.  We can think of no plausible explanation for
why these systems should be absent and thus conclude that the observed
period minimum at $\sim78$ minutes represents the true period minimum for
CVs. If this is indeed the case, we must then explain why SDSS J1507
has an orbital period significantly below 78 minutes.

Calculations show that CVs containing evolved secondary stars, in which
a significant fraction of hydrogen in the core is processed prior to
mass transfer, can reach orbital periods significantly below the
period minimum \citep[e.g][]{thorstensen02,podsiadlowski03}. Indeed,
there is good reason to suspect that the two other CVs with periods
below the observed period minimum contain evolved secondary stars.  EI
Psc has a secondary which is unusually hot (K4$\pm$2 -
\citealt{thorstensen02}) and unusually massive ($\sim0.12 M_{\odot}$ -
\citealt{skillman02}) for its orbital period, and the secondary star
in EI Psc has a very large N/C abundance \citep{gaensicke03}. The mass
and spectral type of the secondary star in EI Psc agree well with an
evolutionary sequence in which a 1.2$M_{\odot}$ star begins mass
transfer close to the end of its hydrogen burning
phase \citep{thorstensen02}. V485 Cen, too, has a mass which is large
for its orbital period ($\gta 0.14 M_{\odot}$ -
\citealt{augusteijn96}), and both stars have He\,{\small I}
$\lambda6678$/H$\alpha$ ratios higher than those typically seen in
 SU UMa systems \citep{augusteijn96,thorstensen02}, suggestive
of an enhanced Helium abundance in the donor.

\begin{figure}
\begin{center}
\includegraphics[scale=0.35,angle=-90,trim=0 30 -0
0]{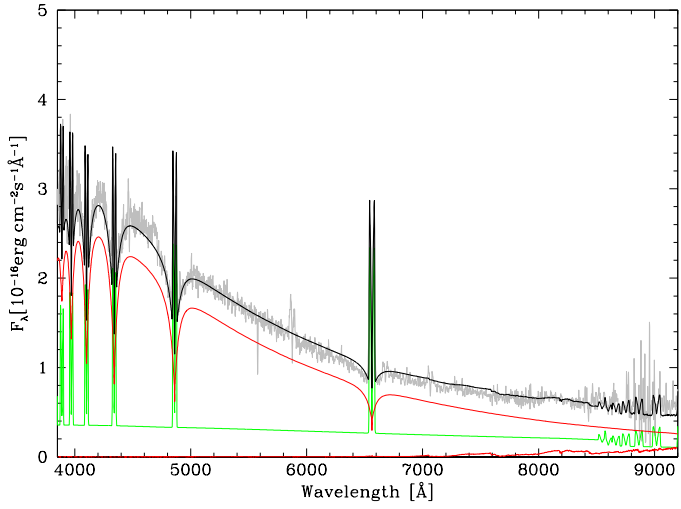}
\caption{Model fits to the SDSS optical spectrum of SDSS J1507. The
light grey curve is the actual spectrum. The white dwarf contribution
is plotted as a solid red line whose flux increases towards the
blue. The accretion disc contribution is a solid green line and the
contribution from an M9V star is plotted as a solid red line whose
flux increases towards the red.}
\label{fig:fit}
\end{center}
\end{figure}

In the case of SDSS J1507, however, an extremely evolved secondary
star cannot be the explanation for the short orbital period. The mass
of the secondary ($M_r = 0.056 \pm 0.001 M_{\odot}$) is a factor of
two below the secondary star masses in EI Psc and V485 Cen.  The
absence of secondary star features in the optical spectrum also
suggests a rather cool secondary star. We modelled the spectrum of
SDSS J1507 using three components following the prescription outlined
in \cite{gaensicke06}. The white dwarf spectrum is computed for the
parameters listed in table~\ref{parameters} using {\small
SYNSPEC/TLUSTY} \citep{hubeny95}. The accretion disc is described in
terms of an isothermal/isobaric slab of hydrogen \citep{gaensicke99}
and the secondary star is represented by M/L dwarf spectral templates
from \cite{beuermann98} and \cite{kirkpatrick00}. The secondary star's
radius is fixed to 0.097$R_{\odot}$ (table~\ref{parameters}). Free
parameters in this model are the temperature, column density and flux
scaling factor of the hydrogen slab as well as a common scaling factor
for the white dwarf/secondary star fluxes. A plausible fit is shown in
figure~\ref{fig:fit}, which implies a spectral type later than M9,
which corresponds to an effective temperature of $T^{eff}_r \lta
2450$K \citep{vrba04}. This compares to a spectral type in EI Psc of
K4$\pm$2 \citep{thorstensen02}.  Also, the He\,{\small I}
$\lambda6678$/H$\alpha$ ratio in SDSS J1507 is 0.07. This is
significantly lower than the ratio of 0.28 measured in EI Psc and
closer to the average ratio amongst SU UMa stars of 0.13
\citep{thorstensen02}. The He\,{\small I} $\lambda6678$/H$\alpha$
ratio depends sensitively on density and temperature as well as
abundance, so this evidence is not conclusive, but it does agree with
the general picture that the donor in SDSSJ1507 is not as evolved as
those in EI Psc and V485 Cen.

\begin{figure}
\begin{center}
\includegraphics[scale=0.45,angle=0,trim=0 0 -0 0]{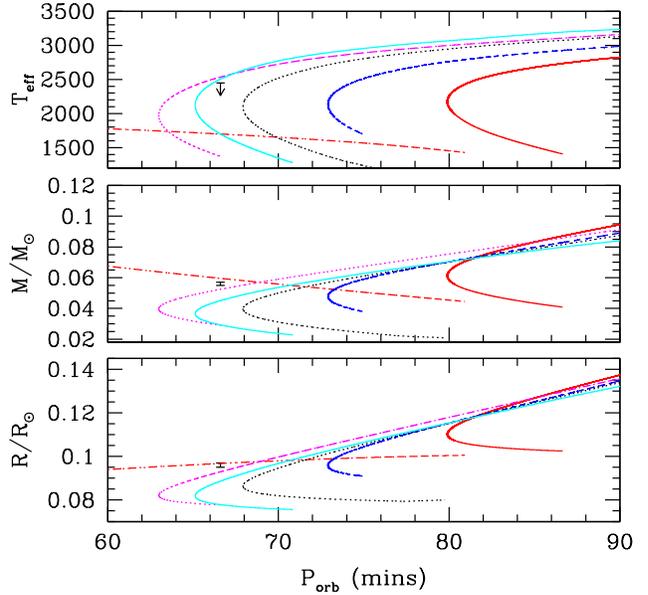} 
\caption{The system parameters of SDSS J1507, compared to the
predictions of some illustrative evolutionary sequences. Solid red
line: Solar hydrogen abundance, $\dot{J}=2\dot{J}_{GR}$. Blue dashed
line: an evolved sequence with hydrogen fraction by mass $H=0.47$,
$\dot{J}=2\dot{J}_{GR}$. Black dashed line: $H=0.3,
\dot{J}=2\dot{J}_{GR}$. Solid cyan line: $H=0.2,
\dot{J}=2\dot{J}_{GR}$. Dashed magenta line: $H=0.3,
\dot{J}=\dot{J}_{GR}$. Red dashed line: sequence with Solar abundance,
an initial donor mass of 0.07$M_{\odot}$ and $\dot{J}=2\dot{J}_{GR}$
where $\dot{J}_{GR}$ is the angular momentum loss rate predicted by
gravitational radiation.}
\label{fig:sequences}
\end{center}
\end{figure}

There is, of course, a continuum of possibilities between the highly
evolved secondaries in EI Psc and V485 Cen, and a secondary with a
normal hydrogen abundance. In order to see whether a moderately
evolved secondary star could be compatible with the observed system
parameters we computed a set of illustrative evolutionary sequences,
based upon the models described in \cite{kolb99}. To ensure our
sequences reproduced the {\em observed} period minimum, we adopted an
arbitrarily enhanced angular momentum loss rate, twice that predicted
by gravitational radiation.  We also included the effects of
distortion of the secondary due to tidal and rotational forces as
quantified by \cite{renvoize02}. Models were calculated for an
initially unevolved secondary star, as well as sequences with initial
hydrogen fractions (by mass) of $H=0.47,0.3$ and 0.2. For $H<0.4$,
sequences were started with initial donor masses below 0.2$M_{\odot}$
and with artificially enhanced abundances. This avoided having to
search for the higher-mass progenitors of these systems (the
progenitors are expected to have an intial mass $\gta 1M_{\odot}$ and
to start mass transfer toward the end of the central hydrogen burning
phase, most likely for a central hydrogen mass fraction $\lta 0.2$
according to the calculations of \cite{baraffe00}).  In order to
quantify the effects of our ad-hoc assumptions about angular momentum
loss on the evolved sequences, the $H=0.3$ sequence was also
calculated with an angular momentum loss rate equal to that predicted
by gravitational radiation.  These sequences are shown in
figure~\ref{fig:sequences}. From this figure it is apparent that both
the sequence with $H=0.2$ and enhanced mass transfer, and the sequence
with $H=0.3$ and ``normal'' mass transfer are broadly consistent with
the observed mass, radius and effective temperature limit of SDSS
J1507.

There are, however, a number of difficulties encountered when we try
to explain the observed properties of SDSS J1507 with evolved
sequences. The evolved sequences do not provide a very good fit to the
observed mass and radius; if we assume a consistent (enhanced) angular
momentum loss rate for both non-evolved and evolved sequences, then
the hydrogen fraction must be lower than $H\lta 0.2$ in order to reach
orbital periods below $\sim 67$ minutes. Such evolved sequences
under-predict the secondary mass.  At a period of 66 mins, the sequence
with $H=0.2$ has a mass of 0.045$M_{\odot}$, which is
inconsistent with the observed mass of $M_r = 0.056 \pm 0.001
M_{\odot}$.  A good fit can be obtained with a sequence with $H=0.3$
and angular momentum loss rates equal to those predicted by
gravitational radiation. Thus, to consistently explain both the system
parameters of SDSS J1507 and the observed period minimum requires that
the angular momentum loss rate be different in evolved and non-evolved
sequences. There is no reason why this might not be so, but it does
imply that the models need some degree of ``tuning'' to consistently
explain the secondary star mass in SDSS J1507 and the observed period
minimum. Additionally, sequences with an evolved secondary star are
only just consistent with the upper limit to the effective temperature
derived above, and fail to explain the slightly lower than average
He\,{\small I} $\lambda6678$/H$\alpha$ ratio in SDSS J1507. It is
therefore difficult, though by no means impossible, to explain the
short period of SDSS J1507 with an evolved secondary star.

A second possibility is that SDSS J1507 formed directly from a
detached white dwarf/brown dwarf binary. \cite{politano04} showed that
CVs forming via this route could have orbital periods as short as 46
minutes.  \cite{maxted06} recently identified the detached white dwarf/brown
dwarf binary WD0137-349, with a white dwarf mass of 0.4$M_{\odot}$ and
a period of 116 minutes, showing that progenitors for such CVs can in
principle survive the common envelope phase (although the white dwarf
mass is very much lower in WD0137-349 than in SDSS
J1507). Furthermore, the observed mass and radius of the secondary
star in SDSS J1507 agree well with the predicted values for a system
in which the initial donor mass was 0.07$M_{\odot}$. Given the
excellent agreement between observed and predicted system parameters,
and the fact that it is the simplest evolutionary scenario
consistent with the data, we believe that it is likely that SDSS
J1507 formed directly from a white dwarf/brown dwarf binary.

\subsection{CVs forming with brown dwarf secondaries}
\label{subsec:direct}

There are now two CVs known which have secondary stars with measured
masses below the substellar limit: SDSS J1507 and SDSS J103533.03+055158.4
\citep{littlefair06}. With an orbital period of 82.1 minutes, SDSS
J1035 is consistent with a CV which has evolved past the period
minimum. As we argue above however, SDSS J1507 has likely formed
directly from a detached white dwarf/brown dwarf binary. It is the
first such system to be identified.

The puzzle is why there are not more. \cite{politano04} found that
15\% of the zero-age CV population should have formed with brown dwarf
secondaries {\em and} have orbital periods below the observed period
minimum. Since the evolution of these objects is slow, we might expect
the present-day CV population to be broadly similar in this respect
\citep{politano04}. Discounting V485 Cen and EI Psc on the basis of
their evolved secondaries, there is now evidence for at least one such
system, out of the 647 CVs with known periods \citep{ritter06}. Could
observational selection effects explain the discrepancy? The catalogue
of \cite{ritter06} is compiled using a wide range of surveys with
varying selection effects; some CVs are identified using variability,
others from their blue colours and others as a result of their large
outbursts \citep{gaensicke05}. However, the mass transfer rates in CVs
forming from brown dwarf secondaries are predicted to be comparable to
those of CVs above the period minimum \citep{kolb99}, of which large
numbers exist in the Ritter catalogue. Furthermore, the presence of
SDSS J1507 amongst the SDSS CV sample suggests that this survey, at
least, is sensitive to these systems. Within the SDSS CV sample there
are currently 90 CVs with known periods; even within this sub-sample
there is clearly a marked absence of CVs below the observed period
minimum ($\sim 1$\% compared to 15\%).  It is therefore unlikely that
selection effects can explain the absence of significant numbers of
CVs below the observed period minimum.

\subsubsection{Common envelope efficiency}
 
One way of reducing the predicted number of CVs forming with brown
dwarf secondaries is to suggest that they will not survive the common
envelope phase. If the orbital energy is insufficient to remove the
common envelope, a merger will occur. \cite{politano07} showed that,
to reconcile the observed and predicted number of CVs forming with
brown dwarf secondaries, the common envelope efficiency parameter,
$\alpha_{ce}$, would have to drop to zero for secondary masses lower
than $\sim 0.1 M_{\odot}$. The existence of SDSS J1507 argues strongly
against this scenario as, by definition, SDSS J1507 must have survived
its common envelope phase. To determine the implications of this for
common envelope efficiency, we adopt equation~1 of \cite{politano07},
and assume the primary in SDSS J1507 was initially a main sequence
star of 3.5--5.5$M_{\odot}$ \citep{dobbie06}. We adopt a value of
$\gamma = 1$ for the dimensionless parameter describing the structure
of the primary, and assume no mass loss occurs prior to the common
envelope phase. We assume that the binary components are co-eval, and
determine a lower limit to $\alpha_{ce}$ by requiring that, at the
end of the common envelope phase, the brown dwarf remains smaller than
its Roche Lobe. For a progenitor mass of 3.5$M_{\odot}$, we find that
$\alpha_{ce} \gta 0.4$, whilst for a progenitor mass of
4.5$M_{\odot}$, we find $\alpha_{ce} \gta 0.9$. For larger progenitor
masses $\alpha_{ce}$ exceeds 1. Significant uncertainties about the
evolutionary state of the primary at the onset of the common envelope
phase means that our results are not unambiguous, and
so these values of $\alpha_{ce}$ should be treated with caution, but
our results do indicate that SDSS J1507 can survive the common
envelope phase for physically plausible common envelope
efficiencies. We note that WD0137-349 provides independent
confirmation that close binaries with brown dwarf secondaries can
survive the common envelope phase \citep{maxted06}. It is therefore
unlikely that the shortage of CVs forming with brown dwarf secondaries
is a result of mergers during the common envelope phase.

\subsubsection{The brown dwarf desert}

Whilst the simulations of \cite{politano07} assume a flat mass ratio
distribution, observations of companions to solar-type stars indicate
a relative scarcity of brown dwarf companions compared with either
planetary mass or stellar companions \citep[e.g.][]{duquennoy91}. This
has been termed the ``brown dwarf desert''. \cite{grether06} found
that nearby binaries with solar-type primaries and orbital separations
$\lta 3$\,AU are $\sim10$ times more likely to have stellar companions
than brown dwarf companions. Around 75\% of the simulated zero-age CVs
forming with brown dwarf secondaries had progenitors consisting of
solar-type primaries with orbital separations less than 3 AU
\citep{politano07}. The dearth of CVs with periods below the period
minimum could therefore be a simple consequence of a shortage of
progenitors for these CVs. Since the observed properties of SDSS J1507
seem to indicate that the shortage of CVs with periods below the
period minimum cannot be explained by observational selection effects
or by mergers during the common envelope phase, we conclude that the
dearth of CVs with periods shorter than the period minimum provides
independent support for the reality of the brown dwarf desert.

\subsection{The white dwarf temperature and ZZ Ceti pulsations}

The observational ZZ Ceti instability strip ranges in effective
temperature from 12270 to 10850K \citep{gianninas05}. Current
theoretical models predict that the ZZ Ceti instability strip should
be pure: all the isolated DA white dwarf stars, with temperatures
within the instability strip limits should pulsate, with a small
dependency on mass \citep{arras06}. With an effective temperature of
$11000\pm500$K, the white dwarf in SDSS J1507 is inside the
conventional ZZ Ceti strip. Lomb-Scargle periodograms reveal no
evidence for ZZ Ceti pulsations, however, despite our rapid photometry
and good signal-noise ratio.

In fact, almost all of the observed pulsating white dwarfs in CVs are
outside the conventional ZZ Ceti strip
\citep{szkody07,szkody02,araujo05}, a result which is probably
explained by the effects of a spread in white dwarf masses and
accretion of small amounts of Helium from the secondary star
\citep{arras06}. These effects may explain the absence of ZZ Ceti
pulsations in SDSS J1507.

\section{Conclusions}

We present high-speed, three-colour photometry of the eclipsing
cataclysmic variable SDSS J150722.30+523039.8.  We measure an orbital
period of 66.61 minutes, placing it below the observed ``period
minimum'' for cataclysmic variables.  By fitting a parameterised model
to the eclipse lightcurves we obtain a mass ratio of $q = 0.0623 \pm
0.0007$, a primary mass of $M_{\rmn{w}} = 0.90\pm0.01 M_{\sun}$ and a
secondary mass and radius of $M_{\rmn{r}} =0.056\pm0.001 M_{\sun}$ and
$R_{\rmn{r}} = 0.096 \pm 0.001 R_{\sun}$, respectively.  The secondary
star in SDSS J1507 has a mass substantially below the hydrogen burning
limit, making it the second confirmed substellar donor in a
cataclysmic variable.

We attempt to explain the short orbital period of SDSS J1507 as a
result of nuclear evolution of the secondary star. We argue that given
the secondary star's low mass, the lack of any visible signature of
the secondary star and the low He\,{\small I} $\lambda6678$/H$\alpha$
line ratio, together with the difficulty in consistently explaining
the secondary star mass in SDSS J1507 and the observed minimum period,
argue against this scenario. Instead, we suggest that SDSS J1507 may have
formed directly from a detached white dwarf/brown dwarf binary. If
this scenario is correct, SDSS J1507 will be the first CV known to
have formed this way. The effective temperature of the secondary, as
well as its surface abundances, ought to be quite different in the two
cases; deep infrared spectroscopy of SDSS J1507 would be highly
desirable.

If SDSS J1507 did indeed form directly from a detached white
dwarf/brown dwarf binary, its observed properties argue against the
scarcity of such systems being explained by observational selection
effects, or mergers during the common envelope phase. The best
remaining explanation is that the progenitors of CVs are themselves
scarce, providing independent evidence for the existence of the brown
dwarf desert.

\section*{\sc Acknowledgements}
TRM acknowledges the support of a PPARC Senior Research
Fellowship. CAW acknowledges the support of a PPARC Postdoctoral
Fellowship.  ULTRACAM and SPL are supported by PPARC grants
PP/D002370/1 and PPA/G/S/2003/00058, respectively. This research has
made use of NASA's Astrophysics Data System Bibliographic
Services. Based on observations made with the William Herschel
Telescope operated on the island of La Palma by the Isaac Newton Group
in the Spanish Observatorio del Roque de los Muchachos of the
Instituto de Astrofisica de Canarias.

\bibliographystyle{mn2e}
\bibliography{abbrev,refs,refs2}

\end{document}